\documentclass[prl,twocolumn,floatfix,superscriptaddress]{revtex4}

         \usepackage{epsf}
         \usepackage{graphicx}
         \usepackage{dcolumn}
         \usepackage{bm}

         \newcommand{\be}{\begin{equation}}
         \newcommand{\ee}{\end{equation}}
         \newcommand{\ba}{\begin{eqnarray}}
         \newcommand{\ea}{\end{eqnarray}}
         \newcommand{\nn}{\nonumber \\}

\begin{document}

         \title{Macroscopic Resonant Tunneling in the Presence of Low Frequency Noise}

         \author{M.~H.~S.~Amin}
         \affiliation{D-Wave Systems Inc., 100-4401 Still Creek Drive,
         Burnaby, B.C., V5C 6G9, Canada}

         \author{Dmitri V.~Averin}
         \affiliation{Department of Physics and Astronomy, Stony Brook
         University, SUNY, Stony Brook, NY 11794-3800 }


         \begin{abstract}

         We develop a theory of macroscopic resonant tunneling of flux in a
         double-well potential in the presence of realistic flux noise with
         significant low-frequency component. The rate of incoherent flux
         tunneling between the wells exhibits resonant peaks, the shape and
         position of which reflect qualitative features of the noise, and can
         thus serve as a diagnostic tool for studying the low-frequency flux
         noise in SQUID qubits. We show, in particular, that the
         noise-induced renormalization of the first resonant peak provides
         direct information on the temperature of the noise source and the
         strength of its quantum component.

         \end{abstract}

         \maketitle

         Superconducting qubits based on different forms of quantum dynamics
         of magnetic flux in SQUIDs continue to demonstrate steady progress --
         see, e.g., \cite{b1,b2,b3}. Nevertheless, the low-frequency flux noise
         in SQUID structures still provides the major obstacle to their further
         development to the level necessary for quantum computing applications.
         Because of this noise, quantum coherence is essentially reduced to the
         states with the same average flux (e.g., to qubit operation at the
         ``optimal point'') \cite{b1,b3,b20}. Only very limited coherence
         exists between the states which differ by their flux values \cite{b2}.
         In superconducting charge qubits, the low-frequency charge noise is
         most probably produced by elementary charges tunneling between
         impurity states localized in the insulator parts of the structure
         \cite{b4}. In contrast to this, it is more difficult to
         construct a possible model of the low-frequency noise of magnetic
         flux. Elementary magnetic moments of nuclei or electrons \cite{b5}
         are small and require very large concentration of impurity states to
         explain the typical value $\delta \Phi \sim 0.1$ m$\Phi_0$ of the
         observed noise \cite{b6,b7}. So the
         origin of the low-frequency flux noise in SQUID structures remains
         not fully understood, making it interesting to probe its properties
         in different experiments. In this work, we show that one of the
         interesting dynamic processes in SQUIDs, macroscopic resonant tunneling
         of flux between the wells of the double-well potential \cite{b10,b11}
         can serve as noise diagnostic tool. In particular, the shape and position
         of the resonant flux tunneling peaks reflects the temperature and the
         strength of quantum component of the low-frequency noise.

         \begin{figure}[t]
         \includegraphics[width=8.7cm]{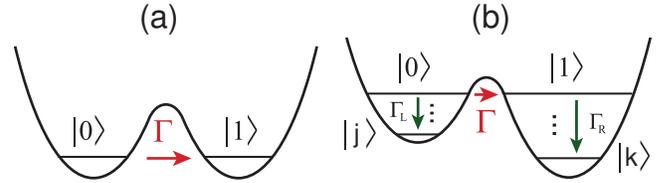}
         \setlength{\unitlength}{1.0in}
         \caption{\label{f1} Schematic energy diagram of the macroscopic
         resonant tunneling in the double-well system between (a)
         two lowest energy states and (b) two excited states in each well. }
         \end{figure}

         Specifically, we consider a SQUID in the regime when its
         flux potential has the double-well form (Fig.~\ref{f1}).
         The extraction of precise parameters of the system, e.g., the
         spectrum $\varepsilon_j$ of the energy levels in the two wells,
         requires its direct numerical simulation. Considerable insight
         into the flux dynamics in this system can still be gained from
         analytical treatment based on the general features of the
         spectrum. The states
         localized in each well are separated by large energy intervals on
         the order of the plasma frequency $\omega_p$, while the tunnel
         coupling $\Delta$ between the states in the opposite wells is
         small ($\hbar=k_B=1$), $\Delta \ll \omega_p$. This means that
         the rate of flux transfer between the
         two wells exhibits resonant peaks as a function of bias between the
         wells \cite{b10} whenever the energy distance $\epsilon$ between a
         pair of levels in the opposite wells becomes small, $\epsilon \ll
         \omega_p$. In the vicinity of such a resonance, the
         influence of the environment on the flux dynamics in this system of
         levels separates into two different effects \cite{b11} even if the
         dissipation is produced by physically one and the same environment.
         The first is ``intrawell'' relaxation that causes the dissipative
         transitions within each well, and is sensitive to the properties of
         environment at large energies $\sim \omega_p$, and the second is
         ``interwell'' relaxation associated with relaxation dynamics within
         the two tunnel-coupled states, and sensitive to the environment
         properties at low energies $\ll \omega_p$. In terms of the coupling
         of flux to the environment, interwell relaxation is dominantly
         affected by fluctuations of the bias $\epsilon$ that shift one well
         of the double-well system as a whole relative to the other. The
         purpose of this work is to extend the previous theories
         \cite{b11,b12} of the resonant flux tunneling to the regime of
         strong interwell relaxation. This regime is made experimentally
         relevant by apparently unavoidable low-frequency flux noise in the
         SQUID structures. The developed approach should also be valid for
         the description of tunneling in other double-well systems besides
         flux qubits.

    We start by calculating the transition rate for tunneling between
    the lowest energy levels $|0\rangle$ and $|1\rangle$ in the ``left"
    and ``right" wells, respectively (Fig.~\ref{f1}a). Such a process
    can be described by a two-state model: $H=H_S+H_B+H_{\rm int}$,
    where $H_B$ is the environment's Hamiltonian and
    \be
    H_S = -{1 \over 2}(\Delta \sigma_x + \epsilon \sigma_z), \qquad
    H_{\rm int} = - {1\over 2} \sigma_z Q \label{Hint}
    \ee
    are the system and interaction Hamiltonians respectively. Here,
    $\sigma_{x,z}$ are the Pauli matrices and $Q$ is an operator acting
    on the environment. We adopt the Gaussian approximation for which
    the dissipative dynamics is characterized completely by the spectral
    properties of the reservoir force $Q$, hence the reservoir
    Hamiltonian $H_B$ does not need to be made explicit. If the
    environment is in equilibrium at temperature $T$, the
    (unsymmetrized) spectral density of the noise $Q$, given by
         \[ S(\omega) = \frac{1}{2\pi} \int_{-\infty}^\infty dt \
         e^{i\omega t}\langle Q(t)Q(0)\rangle \, , \]
    where $ \langle \, ...\, \rangle \equiv \text{Tr}_B \{ \rho \, ...
    \}$, can be expressed in the usual way in the basis of energy
    eigenstates $|n \rangle$ of $H_B$ with eigenvalues ${\cal E}_n$ and
    equilibrium probability density $\rho_n$
         \begin{equation}
         S(\omega) = \sum_{nm} \rho_n |\langle n|Q|m \rangle|^2
         \delta({\cal E}_n-{\cal E}_m+\omega) \, . \label{e5} \end{equation}

    We find the rate $\Gamma$ of flux tunneling between the resonant
    states $|0\rangle$ and $|1\rangle$ in the incoherent regime $\Delta
    \ll W$, where $W$ is the noise-induced resonance width. In this case,
    $\Gamma$ can be calculated in the lowest-order perturbation theory
    in the tunneling term $V = \Delta \sigma_x/2$ in the Hamiltonian
    (\ref{Hint}). We assume that the width $W$ is still small on the scale
    of the plasma energy $\omega_p$, so that the system dynamics can be
    reduced to the two resonant levels even in the presence of noise. This
    condition is satisfied in typical experiments, where $W$ is on the
    order of a few GHz, while $\omega_p \simeq 30$ GHz \cite{b10,Harris07}.
    The lowest-order transition rate between the states $|i\rangle$ and
    $|f\rangle$ can be obtained using Fermi Golden rule:
    \ba \Gamma_{i \to f} &=& 2\pi |\langle i|V|f\rangle|^2 \delta (E_i -
    E_f)\nn &=& \int_{-\infty}^\infty dte^{i(E_i - E_f)t} |\langle
    i|V|f\rangle|^2 \nn &=& 2 \text{Re} \int_0^\infty dt \langle
    i,t|V|f,t\rangle \langle f,0|V|i,0\rangle \, . \label{Gammaif} \ea
    where $E_{i,f}$ are the eigenvalues of the {\em unperturbed} total
    (system + environment) Hamiltonian for states $|i\rangle,|f\rangle$.
    In our case, these states are the systems resonant flux states
    $|0\rangle$ and $|1\rangle$ plus the states of the environment at
    the beginning and the end of the transition. The time evolution of
    these states can be written as ($\alpha=i,f$): \be |\alpha,t\rangle
    =  U(t) e^{i\epsilon \sigma_z t/2} |\alpha,0\rangle. \label{Ut} \ee
    Here, $U(t) = {\cal T}\exp \{(i/2)\, \sigma_z \int_{-\infty}^t
    Q(\tau) d\tau \}$ is the evolution operator in the interaction
    representation, with ${\cal T}$ denoting the time ordering and
    $Q(t)=e^{iH_B t}Qe^{-iH_B t}$. To avoid description of the process
    of ``switching on'' of in general strong interaction with the
    low-frequency environment, we extended the starting time of the
    evolution $U(t)$ to $-\infty$. Substituting (\ref{Ut}) into
    (\ref{Gammaif}), we find $\Gamma_{0 \rightarrow 1} (\epsilon) =
    \Gamma_{1 \rightarrow 0} (-\epsilon) \equiv \Gamma(\epsilon)$, with
         \begin{equation}
         \Gamma(\epsilon) = {\Delta^2\over 2} \text{Re} \int_0^{\infty} dt
         e^{i\epsilon t} \langle U_-^{\dagger} (t) U_+(t)
         U_+^{\dagger} (0) U_- (0) \rangle  , \label{e8}
         \end{equation}
         where $U_\pm = {\cal T}\exp \{\pm(i/2)\int_{-\infty}^t
    Q(\tau) d\tau \}$. Here, we have summed the rates $\Gamma_{i \to f}$
    over the initial (with equilibrium density matrix $\rho$) and final
    states of the environment. The correlator in Eq.~(\ref{e8}) can be
    calculated in
         the Gaussian approximation by expanding each of the operators $U_\pm$ up
         to the second order in $Q$, averaging the result and
         exponentiating it back. This gives:
         \begin{eqnarray}
         \langle U_-^{\dagger} (t) U_+(t) U_+^{\dagger} (0) U_-
         (0)\rangle = \;\;\;\;\;\;\; \nonumber \\ = \exp \left\{ \int_0^{t}
         d\tau \int_{-\infty}^{0} d\tau' \langle Q(\tau) Q(\tau') \rangle \right\}
         . \nonumber \end{eqnarray}
         Expressing the noise correlator in this equation through the
         spectral density (\ref{e5}), we reduce Eq.~(\ref{e8}) to
         \begin{equation}
         \Gamma(\epsilon) = {\Delta^2 \over 2}\text{Re} \int_0^\infty dt
         e^{i\epsilon t} \exp \left\{ \int d \omega
         S(\omega) \frac{e^{-i\omega t}{-}1}{\omega ^2}  \right\} .
         \label{e9} \end{equation}

         The integral over time in Eq.~(\ref{e9}) converges on the scale
         $W^{-1}$ which defines the width $W$ of the resonance. The
         line-shape of the resonance is determined therefore
         by the noise frequencies $\omega <W$. If $S(\omega)$ is essentially
         constant in this frequency range, $S(\omega)=S(0)$, i.e., if the
         noise is sufficiently broad-band and/or weak, the tunneling rate
         (\ref{e9}) has Lorentzian line-shape:
         \begin{equation}
         \Gamma = {1 \over 2}{\Delta^2 W \over \epsilon^2+W^2},
         \qquad W=\pi S(0) \, . \label{e11}
         \end{equation}

         In general, Eq.~(\ref{e9}) can lead to line-shapes that are not
         Lorentzian. If the cut-off frequency $\omega_c$ of the low-frequency
         part of the noise satisfies
         the condition $\omega_c \ll W$ (i.e., the noise is strong and has
         low-frequency), then one can expand the exponent $e^{-i\omega t}$ up
         to the second order in $\omega t$ to obtain:
         \begin{eqnarray}
         \Gamma (\epsilon) = \sqrt{\pi \over 8}{\Delta^2 \over W} \exp
         \left\{-{(\epsilon - \epsilon_p)^2 \over 2W^2}\right\}, \ \
         \label{e15} \\
         W^2 = \int d\omega S(\omega), \qquad  \epsilon_p = {\cal P}\int d\omega
         {S(\omega) \over \omega}. \label{e16}
         \end{eqnarray}
         We assume that both integrals in (\ref{e16}) are finite, with the
         second integral being understood as a principle value. We see that,
         as usual, the noise with a significant low-frequency part yields a
         Gaussian line-shape, in contrast to the Lorentzian line-shape
         (\ref{e11}) in the case of the broad-band noise. Equation
         (\ref{e15}) shows also that, in general, the low-frequency noise not
         only broadens the resonance but also shifts it to the non-vanishing
         bias $\epsilon \simeq \epsilon_p$. This means that effectively the
         first resonant peak splits in two, described by $\Gamma (\epsilon)$
         and $\Gamma(-\epsilon)$, for the two different directions of tunneling.

         If the environmental source of the low-frequency noise is in
         equilibrium at temperature $T$, then symmetric
         and antisymmetric (in frequency) parts of the noise intensity,
         $S_\pm(\omega) {=} [S(\omega)\pm S(-\omega)]/2$, are
         related by the fluctuation-dissipation theorem:
         $S_-(\omega)= S_+(\omega) \tanh (\omega/2T)$. The antisymmetric
         noise gives also the dissipative part $\sigma (\omega,T)$ of the
         linear response (e.g., conductance) of the environment generating
         the noise: $S_-(\omega)=\omega \sigma (\omega,T)$. Since in (\ref{e16})
         the width $W$ and shift $\epsilon_p$ of the Gaussian tunneling peak
         are determined, respectively, by symmetric and
         antisymmetric parts of the noise:
         $W^2 = \int d\omega S_+(\omega)$ and
         $\epsilon_p = \int d\omega S_-(\omega)/\omega $,
    the fluctuation-dissipation theorem relates $W$ and $\epsilon_p$.
    This relation is particularly simple in the case of low-frequency
    noise, when all the relevant frequencies, $\omega \leq \omega_c$,
    are small on the
    scale of temperature $T$: $\omega \ll T$. For typical cut-off
    frequencies on the order GHz, this condition holds even at
    sub-Kelvin temperatures characteristic for experiments with
    solid-state qubits (1 GHz $\simeq$ 50 mK). In this case, $\tanh
    (\omega/2T)\simeq \omega/2T$, yielding
    \begin{equation} W^2 = 2T  \epsilon_p \, . \label{e17} \end{equation}
    Equation (\ref{e17}) is one of the main conclusions of the theory
    developed in this work. It shows that, qualitatively, non-vanishing
    splitting $2\epsilon_p$ of the two parts of the first resonant peak
    is a manifestation of the finite quantum component of the
    low-frequency flux noise. In the regime of classical noise (which is
    valid, due to the fluctuation-dissipation theorem, only as a
    high-temperature approximation), the spectrum $S(\omega)$ is
    symmetric in $\omega$ and $\epsilon_p=0$. Quantitatively, this
    classical limit is reached at $T\gg W$, when $\epsilon_p \ll W$ and
    the two peaks merge.

    The above method of the calculation of the tunneling rates can be
    generalized to the case of resonant tunneling between two higher
    energy levels in the wells. In that case, in addition to the
    interwell relaxation process, we will also have intrawell
    relaxation. We still use $|0\rangle$ and $|1\rangle$ to denote the
    two resonant flux states, which can now be excited states in their
    corresponding wells (see Fig.~\ref{f1}b). In addition to
    Hamiltonians (\ref{Hint}), we need to add to the total Hamiltonian
    two other terms: $H_0 =\sum_{k\neq 0,1} \varepsilon_k |k\rangle
    \langle k|$ that describes the energy of the flux energy eigenstates
    $|k\rangle$ with $k\neq 0,1$, and
    \ba && H_r =\sum_{k\neq k'} \phi_{kk'} (|k\rangle \langle
    k'|+|k'\rangle \langle k|) Q . \label{e3} \ea
    that produces the intrawell relaxation. The sum in $H_r$ goes over
    the pairs of states $|k\rangle$ and $|k'\rangle$ that belong to the
    {\em same} well and $\phi_{kk'}$ are the coupling matrix elements
    between these states that drive the relaxation transitions.

    We make two main assumptions about the properties of decoherence.
    First is that the high-frequency component of the noise is
    sufficiently weak, so that the intrawell relaxation rates in the two
    wells are small on the scale of $\omega_p$, $\Gamma_{L,R} \ll
    \omega_p$, so that it makes sense to discuss resonances in the
    tunneling rate. The relaxation rates can still be large on the scale
    of tunneling strength $\Delta$. Under a natural assumption $T\ll
    \omega_p$, effects of the weak (on the scale of $\omega_p$)
    intrawell relaxation depend only on the spectral components of the
    $\omega \simeq \omega_p$ that enter into the relaxation rates
    $\Gamma_{L,R}$, and we can characterize them directly by these
    rates.

    We find the rate $\Gamma$ of flux tunneling between the resonant
    states $|0\rangle$ and $|1\rangle$ the same way as before. Since the
    environmental modes that contribute to the interwell and intrawell
    relaxations are from different parts of the environment spectrum and
    the coupling Hamiltonians $H_{\rm int}$ and $H_r$ commute (the
    matrix elements $\phi_{kk'}$ are non-vanishing only between the
    states within the same well), Eq.~(\ref{Ut}) can be written as
    direct combination of the evolution due to the high-frequency noise
    $\tilde{U}(t) = {\cal T} \exp \{ - i\int_0^t H_r(\tau) d\tau \}$ and
    the low-frequency noise $U(t)$, as defined below (\ref{Ut}):
    \[ |\alpha,t\rangle =  \tilde{U}(t) U(t) e^{i\epsilon \sigma_z t}
    |\alpha,0\rangle, \]
    where $H_r (t)=e^{i(H_B+H_0) t}H_r e^{-i(H_B+H_0)t}$. Since the
    relaxation is assumed weak on the scale of the plasma frequency,
    such a trace over the high-frequency part of the environment can be
    done in the relaxation approximation, when the relaxation dynamics
    is determined by the transition of the lowest (second) order in
    $H_r$ that are characterized by the relaxation rates $\Gamma_{L,R}$.
    Then,
         \begin{equation}
         \mbox{Tr}_B \{ \rho \langle j |\tilde{U}^{\dagger} (t) |j \rangle
         \langle j'|\tilde{U}(t) |j'\rangle \} = e^{-\gamma t}, \label{e7}
         \end{equation}
    where $j, j'$ denote the two opposite resonant flux states coupled
    by tunneling: $j, j'\in \{0,1\}$, $j\neq j'$, and $\gamma \equiv
    (\Gamma_L + \Gamma_R )/2$. Equation (\ref{e7}) is written in the
    form convenient for the calculation of the tunneling rate in
    Eq.~(\ref{Gammaif}). It can be understood more directly if one
    notices that it gives the time evolution of the off-diagonal element
    $|j \rangle \langle j'|$ of the flux density matrix averaged over
    the environment. According to the standard theory of weak relaxation
    (see, e.g., \cite{blum}), such a matrix element decays with the rate
    equal to the half-sum of all the relaxation rates out of the states
    $|j \rangle$ and $|j' \rangle$ - cf. Eq.~(\ref{e7}). After the
    high-frequency part of the environment is traced out, expression for
    the flux tunneling rate takes the form
         \begin{equation}
         \Gamma(\epsilon) = {\Delta^2 \over 2}\text{Re} \int_0^\infty dt
         e^{(i\epsilon-\gamma) t} \exp \left\{ \int d \omega
         S(\omega) \frac{e^{-i\omega t}{-}1}{\omega ^2}  \right\} .
         \label{Gamma} \end{equation}

         This equation describes higher resonant peaks, when the flux tunnels
         from the ground state in the initial (e.g., left) well into an
         excited state of the target (right) well. In this case, $\gamma=
         \Gamma_R/2$, and the line-shape of
         the resonance (\ref{Gamma}) is determined by the combined action
         of relaxation broadening and the low-frequency noise. If the noise
         is strong, using the same approximation as for the first
         resonance, we get:
         \begin{equation}
         \Gamma (\epsilon) =  \sqrt{\pi \over 8}{\Delta^2 \over W} \text{Re}
         \left[w\left({\epsilon - \epsilon_p +i\gamma \over \sqrt{2}W}
         \right) \right], \label{e18} \end{equation}
         where the parameters are given as before by Eq.~(\ref{e16}), and
         $w(x)$ is the complex error function defined as
         \[ w(x) =e^{-x^2}[1-\text{erf}(-ix)] = e^{-x^2}{2\over\sqrt{\pi}}
         \int_{-ix}^\infty e^{-t^2}dt \, . \] Equation (\ref{e18}) describes
         the transition between the Gaussian and Lorentzian line-shapes in
         the noise- ($W\gg \gamma$) and the relaxation-dominated ($W\ll
         \gamma$) regimes, respectively. Independently of this, if the noise
         is quantum and $\epsilon_p \neq 0$, the higher resonant peaks are
         also shifted by $\epsilon_p$ away from their positions determined by
         the energy levels in the well that follow from the Schr{\"o}dinger
         equation for the flux dynamics in the absence of environment.

         Finally, we consider the microwave-activated tunneling between
         the two wells. For the simplest case where the
         tunneling occurs between the two lowest energy levels of each
         well, the process can be described by the Hamiltonian
         $ H_{rf} =- \epsilon \sigma_z + \hat V \cos \Omega t, $
         where $\hat V$ is an operator that couples microwaves to the flux
         dynamics. Unitary transformation $U=e^{-i(\Omega t/2) \sigma_z}$
         changes this Hamiltonian into $\tilde{H}_{rf} = UH_SU^\dagger +
         i\dot{U}U^\dagger$, and gives in the rotating wave approximation:
         %
         $\tilde{H}_{rf}= -{1 \over 2} (\tilde{\epsilon} \sigma_z +\tilde{V}
         \sigma_x)$,
         %
         where $\tilde{\epsilon} =\epsilon-\Omega \ll \Omega$ is the detuning
         of microwave radiation and $\tilde{V}=\langle 0|\hat V|1\rangle$. In the
         absence of the environment, the system would undergo Rabi
         oscillations with the Rabi frequency $\Omega_R=\sqrt{
         \tilde{\epsilon}^2+|\tilde{V}|^2}$. If coupling to the environment is
         strong enough so that $\tilde{V}\ll W$, then the tunneling becomes
         incoherent with the rate given by (\ref{e15}), but with
         $\epsilon$ and $\Delta$ replaced by $\tilde{\epsilon}$ and
         $|\tilde{V}|$, respectively.

         In another interesting regime, the microwave excites
         the system from the ground state $|k\rangle$ to an excited
         state $|0\rangle$ in the first well, and it tunnels then
         to state $|1\rangle$ in the target well
         (Fig.~\ref{f1}). We make use of the same assumption of
         small tunnel coupling $\Delta$ that can be treated as perturbation.
         In addition to previously used condition $\Delta \ll W$, this assumption
         implies also that the tunneling rate $\Gamma(\epsilon)$ is small
         compared to the rates of excitation/relaxation processes within each
         well. The balance between excitation and relaxation ($\Gamma_L$) in the
         well establishes the
         stationary occupation $p_0$ of the state $|0\rangle$. The
         standard solution for the relaxation dynamics of the density matrix
         of the system (see, e.g., \cite{b11}) gives
         $p_0= |\tilde{V}|^2 /(\tilde{\epsilon}^2+2 |\tilde{V}|^2 +
         \Gamma_L^2/4)$.
         Here, again, $\tilde{\epsilon}=\epsilon_0-\epsilon_k-\Omega$ is
         the detuning energy, and $\tilde{V}=\langle 0|\hat V|k\rangle$.
         At this stage, one can repeat the same steps in the calculation of
         the incoherent tunneling rate between the states $|0\rangle$ and
         $|1\rangle$ as above, and obtain the rate
         $
         \tilde{\Gamma} =p_0 \Gamma ,
         $
         where $\Gamma$ is given by (\ref{e18}), but now with the
         relaxation broadening determined by the relaxation in both wells,
         $\gamma = (\Gamma_L + \Gamma_R )/2$. Qualitatively, as before,
         the shape of the resonant peak of $\tilde{\Gamma}$ is determined by
         a convolution of the Lorentzian broadening due to intrawell
         relaxation and Gaussian broadening due to interwell noise.

         In conclusion, we have developed the theory of macroscopic resonant
         tunneling in flux qubits in the presence of Gaussian low-frequency
         flux noise. The tunneling rate is given in general by Eq.~(\ref{Gamma}).
         In the case of strong noise and tunneling between the two lowest
         energy levels in each well, the resonant peaks have Gaussian shape
         (\ref{e15}). If the noise source is in equilibrium at temperature
         $T$, there is a fundamental relation
         (\ref{e17}) between the width $W$ of the resonant peaks and shift
         $\epsilon_p$ of their position in energy. The shift $\epsilon_p$
         provides direct measure of the strength of the quantum component in
         the flux noise, and leads to splitting of the first resonant peak of
         flux tunneling. Some of the predictions of the present work has been
         already tested in experiment \cite{Harris07}, which shows, in
         particular, that Eq.~(\ref{e17}) is a convenient tool to determine
         whether the low-frequency flux noise is produced by an equilibrium
         quantum source. More experiments, especially with microwave
         excitations, can be performed to further test our theory.

         The authors would like to thank A.J.~Berkley, S.~Han, R.~Harris,
         J.~Johansson, M.W.~Johnson, K.K. Likharev, J.E. Lukens, G.~Rose,
         and V.K. Semenov for useful discussions. D.V.A. was supported in
         part by the NSA under ARO contract W911NF-06-1-217.

         \end{document}